\documentclass[11pt, a4paper]{article}

\usepackage[utf8]{inputenc}
\usepackage{amsmath}
\usepackage{graphicx}
\usepackage{url}
\usepackage[margin=1in]{geometry}
\usepackage{hyperref}
\hypersetup{
    colorlinks=true,
    linkcolor=blue,
    filecolor=magenta,      
    urlcolor=cyan,
}
\usepackage{listings}
\usepackage{xcolor}

\definecolor{codegreen}{rgb}{0,0.6,0}
\definecolor{codegray}{rgb}{0.5,0.5,0.5}
\definecolor{codepurple}{rgb}{0.58,0,0.82}
\definecolor{backcolour}{rgb}{0.95,0.95,0.92}

\lstdefinelanguage{YAML}{
  keywords={true,false,null,on,off},
  keywordstyle=\color{blue},
  basicstyle=\ttfamily\footnotesize,
  ndkeywords={name, on, jobs, runs-on, steps, uses, with, env, run, permissions, contents, id-token, schedule, cron, workflow_dispatch, secrets},
  ndkeywordstyle=\color{red},
  comment=[l]{\#},
  commentstyle=\color{codegreen},
  stringstyle=\color{codepurple},
  string=[s]{'}{'},
  string=[s]{"}{"},
  morestring=[b]',
  morestring=[b]",
  sensitive=true,
  morecomment=[l]{\#},
}

\lstdefinestyle{pythonstyle}{
    backgroundcolor=\color{backcolour},   
    commentstyle=\color{codegreen},
    keywordstyle=\color{magenta},
    numberstyle=\tiny\color{codegray},
    stringstyle=\color{codepurple},
    basicstyle=\ttfamily\footnotesize,
    breakatwhitespace=false,         
    breaklines=true,                 
    captionpos=b,                    
    keepspaces=true,                 
    numbers=left,                    
    numbersep=5pt,                  
    showspaces=false,                
    showstringspaces=false,
    showtabs=false,                  
    tabsize=2,
    language=Python
}

\title{\textbf{A Serverless Architecture for Real-Time Stock Analysis using Large Language Models: An Iterative Development and Debugging Case Study}}
\author{
  Taniv Ashraf \\
  \texttt{tanivashraf@gmail.com} \\
  \url{https://www.tanivashraf.com} \\
  \textit{Independent Researcher}
}
\date{\today}

\begin{document}

\maketitle

\begin{abstract}
The advent of powerful, accessible Large Language Models (LLMs) like Google's Gemini presents new opportunities for democratizing financial data analysis. This paper documents the design, implementation, and iterative debugging of a novel, serverless system for real-time stock analysis. The system leverages the Gemini API for qualitative assessment, automates data ingestion and processing via GitHub Actions, and presents the findings through a decoupled, static frontend. We detail the architectural evolution of the system, from initial concepts to a robust, event-driven pipeline, highlighting the practical challenges encountered during deployment. A significant portion of this paper is dedicated to a case study on the debugging process, covering common software errors, platform-specific permission issues, and rare, environment-level platform bugs. The final architecture operates at a near-zero cost, demonstrating a viable model for individuals to build sophisticated AI-powered financial tools. The operational application is publicly accessible, and the complete source code is available for review. We conclude by discussing the role of LLMs in financial analysis, the importance of robust debugging methodologies, and the emerging paradigm of human-AI collaboration in software development.
\end{abstract}


\section{Introduction}

The application of Artificial Intelligence to financial markets has been a subject of intense research for decades, traditionally focusing on quantitative models like LSTMs and ARIMA for time-series forecasting \cite{fischer2018}. A recent comprehensive review by Al-Amin (2024) encapsulates the significant progress made using machine learning and deep learning for stock prediction \cite{al2024comprehensive}. However, the recent emergence of highly capable Large Language Models (LLMs) has opened a new frontier for qualitative analysis, allowing for the interpretation of news, sentiment, and complex data patterns in a human-like manner \cite{google2023gemini}. While major financial institutions have the resources to build proprietary AI systems, the accessibility of APIs for models like Google's Gemini provides an unprecedented opportunity for individual researchers and developers.

This project was initiated to explore a central question: \textit{Can a cost-effective, real-time, and fully automated stock analysis system be built and deployed by an individual using publicly available, serverless tools?} The goal was not to create a "get-rich-quick" trading bot, but to develop a framework that could systematically fetch market data, generate qualitative AI-driven insights, and present them in a user-friendly interface. This aligns with modern research trends that seek to combine quantitative data with qualitative, NLP-driven analysis for more holistic financial forecasting \cite{aurito2022}. Similarly, research by Monem et al. has demonstrated the value of integrating sentiment analysis with deep learning models to predict stock price movements, further validating the hybrid approach this paper explores \cite{monem2021}.

This paper documents the complete journey of this project, from conception to a fully operational system. Crucially, it provides a transparent account of the iterative development process, emphasizing the real-world challenges and debugging strategies employed. We believe this case study in iterative problem-solving is as valuable as the final architecture itself, offering a practical guide for others navigating the complexities of modern cloud and AI deployment.

\paragraph{Author's Note on Methodology and Human-AI Collaboration:} This paper and the system it describes were produced through a novel collaborative workflow. The human author (Taniv Ashraf), leveraging his experience, served as the architect, strategist, and project manager. He provided the high-level goals, defined the sequence of tasks, diagnosed logical errors, and specified the desired features and fixes. A Large Language Model (AI) served as the primary executor and tool, responsible for generating the code, drafting the text, and implementing the solutions as directed. This work is therefore a direct product of human-guided AI execution, a paradigm where the human's role shifts from line-by-line coding to strategic oversight and intelligent prompting, while the AI handles the bulk of the implementation.

\section{System Architecture and Methodology}

The system's final architecture is a serverless, event-driven pipeline that maximizes reliability and minimizes cost.

\subsection{Components}
\begin{itemize}
    \item \textbf{Data Source:} The \texttt{yfinance} library \cite{yfinance} for daily end-of-day stock data and the \texttt{NewsAPI} for recent news headlines.
    \item \textbf{Backend Logic:} A Python script (\texttt{generate\_predictions.py}) serves as the core of the backend.
    \item \textbf{AI Analysis Engine:} Google's Gemini Pro model, accessed via its REST API. The model is prompted to return a structured JSON object.
    \item \textbf{Automation \& Hosting:} GitHub Actions \cite{githubactions} is used as a serverless cron job scheduler. The same GitHub repository hosts the code and stores the output \texttt{predictions.json} file.
    \item \textbf{Frontend:} A static website built with HTML, CSS, and vanilla JavaScript, hosted on a custom domain.
\end{itemize}

\subsection{Data Flow \& Logic}
The system operates on a daily cycle:
\begin{enumerate}
    \item A \texttt{cron} schedule in a GitHub Actions workflow triggers the job.
    \item The Python script runs, fetching fresh stock and news data.
    \item The script reads the \texttt{predictions.json} file from the previous day's run to perform an accuracy check.
    \item A detailed prompt is sent to the Gemini API.
    \item The script parses the AI's response and combines it with current price data and accuracy metrics.
    \item The script overwrites the \texttt{predictions.json} file in the repository. A bot user commits and pushes this change.
    \item The live website dynamically fetches this updated JSON file to display the latest analysis.
\end{enumerate}

\section{The Debugging Journey: A Case Study}

The path to a stable deployment was non-linear. This section details the sequence of critical bugs encountered and the strategies used to resolve them, which we believe is a valuable contribution for practitioners.

\subsection{Initial Bug: Data Serialization Failure}
\begin{itemize}
    \item \textbf{Symptom:} The workflow failed during the Python script execution with the error \texttt{TypeError: Object of type Series is not JSON serializable}.
    \item \textbf{Analysis:} This classic error occurs when attempting to serialize a non-standard data type. The \texttt{yfinance} library, built on pandas, returns price data as \texttt{pandas.Series} objects. While this is efficient for data analysis, the standard Python \texttt{json} library does not have a default method for converting this object to text.
    \item \textbf{Resolution:} The solution was to explicitly cast the pandas object to a standard Python data type before it was added to the dictionary destined for JSON serialization. This was a one-line code change that highlights the critical need for data type awareness at the boundaries between different libraries.
\end{itemize}

\subsection{Platform Bug 1: Repository Write Permissions}
\begin{itemize}
    \item \textbf{Symptom:} The Python script executed successfully, but the workflow failed on the final \texttt{git push} command with a \texttt{403 Forbidden} error and the message \texttt{Permission to ... denied to github-actions[bot]}.
    \item \textbf{Analysis:} For security, the default \texttt{GITHUB\_TOKEN} provided to a workflow has read-only access to the repository's contents. It cannot push changes.
    \item \textbf{Resolution:} The solution was to modify the workflow's \texttt{.yml} file to explicitly grant the necessary permissions. This allows the job to write back to its own repository, a common requirement for workflows that generate artifacts.
\end{itemize}

\subsection{Platform Bug 2: The ``Ghost'' Action}
This was the most challenging bug, representing a failure of the platform environment rather than the user's code.
\begin{itemize}
    \item \textbf{Symptom:} The workflow began failing at the very first step, ``Set up job,'' with the error \texttt{Unable to resolve action ... action not found}. This meant the GitHub runner could not download a standard, public action from the marketplace.
    \item \textbf{Analysis \& Iterative Debugging:}
    \begin{enumerate}
        \item Initial attempts involved verifying the \texttt{.yml} syntax and trying different versions of the action. The error persisted.
        \item The runner environment was changed from \texttt{runs-on: ubuntu-latest} to \texttt{runs-on: ubuntu-22.04}. The error persisted.
        \item Account settings were verified to confirm no organization-level security policies were blocking third-party actions.
        \item A minimal test case workflow was created that did nothing but attempt to \texttt{use} the problematic action. It failed with the same error, definitively isolating the problem to the platform's ability to resolve actions for this specific repository.
    \end{enumerate}
    \item \textbf{Resolution:} Having exhausted all logical configuration fixes, we concluded the issue was a platform-level bug. **A new, blank repository was created, and the exact same code and secrets were migrated. The workflow succeeded on the first attempt.** This unconventional solution underscores a critical problem-solving principle: when a system's behavior contradicts its configuration, the environment itself may be the source of the bug, and recreating it is a valid resolution strategy.
\end{itemize}

\section{Code Evolution Snippets}
This section provides concrete examples of the code changes made to resolve the aforementioned bugs.

\subsection{Fixing the Data Serialization Bug}
The original code in \texttt{generate\_predictions.py} caused a \texttt{TypeError}.
\begin{lstlisting}[language=Python, caption={Original problematic code}]
# This line fails because stock_data['Close'].iloc[-1] is a pandas.Series
prediction_record = {
    'symbol': symbol,
    'current_price': round(stock_data['Close'].iloc[-1], 2), 
    **analysis
}
\end{lstlisting}
The fix involved an explicit cast to a standard Python \texttt{float}.
\begin{lstlisting}[language=Python, caption={Corrected code with type casting}]
# FIX: Explicitly cast the value to a standard float
prediction_record = {
    'symbol': symbol,
    'current_price': round(float(stock_data['Close'].iloc[-1]), 2),
    **analysis
}
\end{lstlisting}

\subsection{Fixing the Git Permission Bug}
The original \texttt{.yml} file lacked the necessary permissions for the job to push changes.
\begin{lstlisting}[language=YAML, caption={Workflow file before permission fix}]
jobs:
  build-and-commit:
    runs-on: ubuntu-22.04
    # No permissions block here
    steps:
      # ...
\end{lstlisting}
The fix was to add a \texttt{permissions} block to grant write access to the repository's contents.
\begin{lstlisting}[language=YAML, caption={Workflow file with corrected permissions}]
jobs:
  build-and-commit:
    runs-on: ubuntu-22.04
    
    # FIX: Grant permissions for the job to write to the repository
    permissions:
      contents: 'write'
      
    steps:
      # ...
\end{lstlisting}

\subsection{Simplifying the Workflow}
After encountering the platform bug, the final, successful workflow was simplified to use GitHub's native secrets, eliminating the complex and error-prone Google Cloud authentication dependency.
\begin{lstlisting}[language=YAML, caption={Final, simplified workflow using GitHub Secrets}]
# ...
- name: Setup Python, Install Dependencies, and Run Script
  # FIX: Get secrets directly from the repository's encrypted secrets
  env:
    GEMINI_API_KEY: ${{ secrets.GEMINI_API_KEY }}
    NEWS_API_KEY: ${{ secrets.NEWS_API_KEY }}
  run: |
    # ... python setup and execution ...
\end{lstlisting}

\section{Conclusion and Future Work}

This project successfully demonstrates that a sophisticated, serverless AI analysis system can be built with minimal cost using modern, publicly available tools. The journey highlights several key takeaways. First, the role of LLMs in finance is most effective when they are treated as \textbf{qualitative reasoning engines}, not as numerical calculators. Second, \textbf{serverless architecture, particularly GitHub Actions, is a transformative tool for individual developers}, removing the need to manage a backend server. Finally, the debugging process underscores that in modern software development, \textbf{understanding the platform and environment is as critical as understanding the code.}

The final, operational system stands as a testament to this iterative, collaborative, and persistent development process. The application can be viewed live, and the complete source code is available for public inspection, ensuring full reproducibility.
\begin{itemize}
    \item \textbf{Live Application:} \url{https://codepen.io/tanivashraf/pen/GgpgxBY}
    \item \textbf{Source Code Repository:} \url{https://github.com/TanivAshraf/ai-stock-analyzer}
\end{itemize}

Future work could involve expanding the data sources to include fundamental analysis (e.g., P/E ratios), performing more granular sentiment analysis on news article content, and backtesting the AI's advice over a longer period to quantitatively assess its directional accuracy and potential profitability.

\bibliographystyle{plain}
\bibliography{references}

\end{document}